\documentclass[twocolumn]{revtex4-1}

\usepackage{graphicx}
\usepackage{amsmath}
\usepackage{amssymb}
\usepackage{amsfonts}
\usepackage{upgreek}
\usepackage{comment}
\usepackage{color}

\begin{document}

\title{Quantum annealing with ultracold atoms in a multimode optical resonator}

\author{Valentin Torggler}
\author{Sebastian Kr\"amer}
\author{Helmut Ritsch}

\affiliation{\small Institut f\"ur Theoretische Physik, Universit{\"a}t Innsbruck, A-6020~Innsbruck, Austria}

\begin{abstract}
A dilutely filled $N$-site optical lattice near zero temperature within a high-$Q$ multimode cavity can be mapped to a spin ensemble with tailorable interactions at all length scales. The effective full site to site interaction matrix can be dynamically controlled by the application of up to $N(N+1)/2$ laser beams of suitable geometry, frequency and power, which allows for the implementation of quantum annealing dynamics relying on the all-to-all effective spin coupling controllable in real time. Via an adiabatic sweep starting from a superfluid initial state one can find the lowest energy stationary state of this system. As the cavity modes are lossy, errors can be amended and the ground state can still be reached even from a finite temperature state via ground state cavity cooling. The physical properties of the final atomic state can be directly and almost non-destructively read off from the cavity output fields. As example we simulate a quantum Hopfield associative memory scheme. 
\end{abstract}

\maketitle

\section{Introduction}
The realization of strong collective coupling between ultracold atoms and the electromagnetic field in a Fabry-P\'erot cavity \cite{ritsch2013cold} opens a unique test ground to study the real time dynamics of quantum phase transitions in open systems of mesoscopic size \cite{black2003observation,slama2007cavity,treutlein2007bose, gupta2007cavity,baumann2010dicke,arnold2012self, kessler2014steering,kollar2015adjustable}. Cavity field mediated interactions induce a variety of self-ordered phases where the particles break the translational symmetry by forming complex spatial patterns \cite{domokos2002collective,ritsch2013cold,keeling2014fermionic,piazza2014umklapp,chen2014superradiance}. In a seminal experiment at ETH the first controllable quantum simulation of the superradiant Dicke phase transition was demonstrated as predicted for the Tavis-Cummings model several decades ago \cite{hepp1973superradiant,baumann2010dicke}. By adding an extra optical lattice in the cavity, the complex phase diagram of a Bose-Hubbard Hamiltonian with tailorable short and infinite range interactions was then experimentally studied in great detail, exhibiting superfluid, insulator and supersolid regions \cite{landig2016quantum}. The experiment shows very good agreement with theoretical models using various approximate numerical methods like dynamical mean field approaches, predicting a supersolid phase region \cite{li2013lattice,bakhtiari2015nonequilibrium}.

In recent work we exhibited that versatility and complexity of the lattice cavity system strongly increase by adding extra pump laser frequencies close to resonance with different cavity modes \cite{kramer2014self}. For classical point particles one finds that the coupled atom-cavity dynamics can be designed as a self-optimizing light collection system with learning and memory capacity \cite{torggler2014adaptive}. Similarly, generalizing the system to fixed multilevel atoms and using degenerate modes, Gopalakrishnan and coworkers previously proposed to simulate a quantum version of the Hopfield model \cite{gopalakrishnan2009emergent,gopalakrishnan2011frustration,gopalakrishnan2012exploring}. Applications to study the physics of a Bose glass were also suggested \cite{habibian2013bose}.

As the scattered light contains information on the atoms' quantum statistical properties, one can perform minimally perturbing observations in real time and use quantum measurement back action and feedback to further control the system \cite{mekhov2009quantum,caballero2015quantum}. First experimental studies of multimode systems were also reported recently \cite{kollar2016supermode}.

\begin{figure}[htp]
\centering
\includegraphics[scale=1]{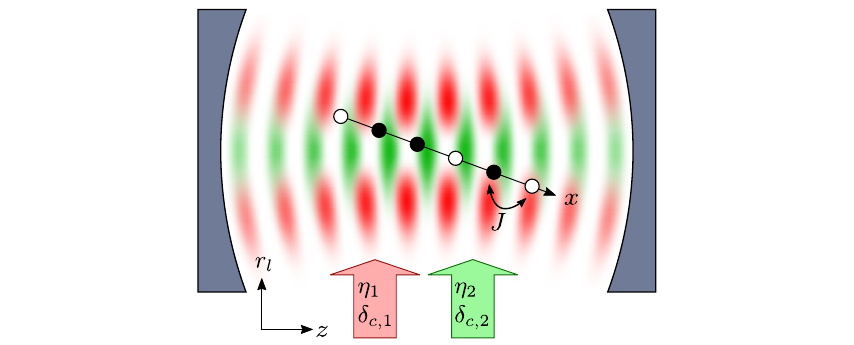}
\caption{A partially filled optical lattice with $N$ sites inside a multimode optical resonator is pumped from the side by several lasers with frequencies close to cavity resonances.}
\label{fig:setup}
\end{figure}

For a single laser frequency the interaction between the atoms induced by a single cavity mode is spatially periodic and infinite range \cite{ritsch2013cold}. In contrast, we show that by help of several pump laser frequencies and tailored illumination geometries, the coupling strengths and light shifts at different sites can be individually modified in such a form as to implement a full connectivity matrix between all lattice sites. In the limit of strong on-site repulsion and low density one gets only zero or one atom per site mimicking a pseudo spin lattice. At least in principle any coupling matrix can be realized using order $N^2$ cavity modes \cite{hauke2015probing}. In contrast to current implementations \cite{johnson2011quantum}, which use minor embedding \cite{choi2008minor}, and alternative architectures \cite{lechner2015quantum}, our approach does not need auxiliary qubits to realize long-range coupling.

As power and frequency of the pump lasers can be externally controlled in real time, we have a natural and straightforward way to implement quantum annealing \cite{kadowaki1998quantum,farhi2001quantum}. One simply slowly increases the strength of the pump lasers in the system to adiabatically reach the ground state of the coupled spin Hamiltonian. As we have a genuine open system, our implementation also suggests a new route towards quantum simulation in a driven-dissipative system as small errors during the sweep process can be amended via cavity ground state cooling \cite{sandner2015self} when we operate the lasers red detuned. This is the more effective the more laser modes we have available for coupling and cooling.

This work is organized as follows: after introducing the general multimode atom-field Hamiltonian and its truncated Bose-Hubbard form, we map it to a coupled spin model in the strong on-site interaction limit in Section \ref{sec:model}. Spin-spin coupling arises from cavity enhanced light scattering and we exhibit how any desired coupling matrix can be found by proper choice of laser parameters in Section \ref{sec:interactionmatrix}. By employing an adiabatic passage described in \ref{sec:quantumannealing} we finally simulate a Hopfield associative memory model via quantum annealing \cite{santra2016exponential} as generic nontrivial example in Section \ref{sec:associativememory}.

\section{Model}
\label{sec:model}
We study a 1D optical lattice with $N_A$ atoms trapped in $N > N_A$ sites, which is placed inside an optical resonator supporting several non-degenerate modes (see Figure \ref{fig:setup}). The atoms are directly illuminated by $M$ lasers with frequencies close to the resonance of the corresponding cavity modes. For sufficient mode spacing, light from each laser is scattered into one specific mode only and scattering between different modes is suppressed. Furthermore, the laser frequencies are far away from any internal atomic resonance which allows for the elimination of the inner atomic degrees of freedom, resulting in an effective Hamiltonian, coupling modes and atomic motion \cite{ritsch2013cold}. In addition to the coherent processes described so far, photons leak out through the mirrors. Note that lattice and cavity orientation can be chosen independently.

The single particle Hamiltonian for an atom with mass $m_\mathrm{A}$ then reads \cite{maschler2008ultracold}
\begin{equation}\label{eq:singleparticlehamiltonian}
\begin{aligned}
H_0 =& \frac{\hat p^2}{2 m_\mathrm{A}} + V_\mathrm{L} \cos^2(k_\mathrm{L} \hat x) - \hbar \sum_{m=1}^M \Delta_{c,m} a_m^\dagger a_m \\
&+ \hbar \sum_{m=1}^M \eta_m (u_{p,m}^*(\hat x) u_{c,m}(\hat x) a_m + \mathrm{h.c.}),
\end{aligned}
\end{equation}
where h.c.\ denotes the Hermitian conjugate. The operators $\hat x$ and $\hat p$ are position and momentum operators along the lattice axis $x$, while $a_m$ ($a_m^\dagger$) denotes the annihilation (creation) operator of a photon in the $m$-th cavity mode. The normalized mode functions of pump and cavity modes are $u_{p,m}(x)$ and $u_{c,m}(x)$, respectively, which are evaluated on the lattice axis. The cavity parameters consist of the effective pump strengths $\eta_m$ and the detuning between pump laser and cavity mode frequency $\Delta_{c,m}$ of the $m$-th mode. The optical lattice of depth $V_\mathrm{L}$ is created by an extra standing wave with wave number $k_\mathrm{L}$. Here we neglect the atomic state dependent dispersive shifts of the cavity modes $N U_{0,m}$ \cite{ritsch2013cold}, which is valid if $|\Delta_{c,m}| \gg N |U_{0,m}|$.

The many-particle Hamiltonian including contact interactions between atoms can be deduced in the framework of second quantization.  In the tight binding limit and neglecting cavity modifications of the tunneling we obtain a generalized intra-cavity Bose-Hubbard Hamiltonian \cite{maschler2005cold, maschler2008ultracold,mekhov2012quantum}
\begin{equation}
\label{eq:model}
\begin{aligned}
H =& H_\mathrm{BH} - \hbar \sum_m \Delta_{c,m} a_m^\dagger a_m\\
& + \hbar \sum_{m} \eta_m \sum_i ((v_m^i)^* a_m + v_m^i a_m^\dagger) \hat n_i
\end{aligned}
\end{equation}
with the standard Bose-Hubbard Hamiltonian
\begin{equation} 
H_\mathrm{BH} = - J \sum_i (b^\dagger_{i+1}b_i + b^\dagger_i b_{i+1}) + \frac{U}{2} \sum_i \hat n_i (\hat n_i - 1).
\end{equation}
Here $b_i$ and $b_i^\dagger$ are bosonic annihilation and creation operators, whereas $\hat n_i = b_i^\dagger b_i$ gives the number of atoms at site $i$. The matrix elements $J$ and $U$ are the nearest neighbor tunneling rate and the on-site repulsion energy \cite{jaksch1998cold}, respectively, which depend on the optical lattice only. The geometry of the modes (i.e.\ the mode functions and laser illumination directions) only enters via the $N$-dimensional coupling amplitude vectors $\boldsymbol{v}_m$ given by
\begin{equation}
v_m^i = \int \mathrm dx \, w^2(x-x_i) u_{p,m}(x) u_{c,m}^*(x),
\end{equation}
where $w(x-x_i)$ is the Wannier function for an atom at site $i$. Thereby we assume that the external optical lattice is much deeper than the potential created by the dynamical cavity field intensity and the pump intensity. In this limit the Wannier functions exclusively depend on the external lattice and are obtained from its Bloch waves in the standard way \cite{kohn1959analytic}.

Additionally to the coherent dynamics treated so far, the cavity fields decay to a steady state with the rates $2 \kappa_m$. If these rates are much larger than the rate of change of the atomic motion $J/\hbar$, the state of the cavity fields instantaneously reacts on an altered atomic state and is thus totally determined by the latter. This can be formally expressed by substituting the field operators by atomic operators
\begin{equation}
\label{eq:fieldeq}
a_m \equiv \eta_m \sum_i v_m^i \hat n_i / (\Delta_{c,m} + i \kappa_m),
\end{equation}
which amounts to an adiabatic elimination of the cavity field operators \cite{nagy2009nonlinear,maschler2008ultracold,habibian2013bose} (see Appendix \ref{app:adiabaticelimination} for details). Note that this a realistic regime: Already for moderately deep lattices $V_\mathrm{L} \sim 10 E_\mathrm{R}$ the matrix elements $J \sim 10^{-2} E_\mathrm{R}$ are much smaller than realistic cavity decay rates $\hbar \kappa_m \gtrsim E_\mathrm{R}$ \cite{baumann2010dicke,wolke2012cavity}, where $E_\mathrm{R} = (\hbar k_\mathrm{L})^2/(2m_\mathrm{A})$ is the recoil energy.


In this so-called bad cavity limit the coherent dynamics is described by an effective atomic Hamiltonian
\begin{equation}
\label{eq:adiabatichamiltonian}
H_\mathrm{ad} = H_\mathrm{BH} - \zeta \sum_{i, j} A_{ij} \hat n_i \hat n_j.
\end{equation}
The interesting part of the physics is encoded in the real and symmetric interaction matrix
\begin{equation}
\label{eq:interactionmatrix}
A = \sum_m (f_m/\zeta) V_m
\end{equation}
with an effective interaction strength $\zeta = \lVert \sum_m f_m V_m \rVert$ and the trace norm $\lVert M \rVert = \mathrm{Tr}(\sqrt{M^\dagger M})$ for some matrix $M$. Thereby each single mode contributes to $A$ with the single mode interaction matrix
 \begin{equation}
 V_m = \mathrm{Re}(\boldsymbol{v}_m \otimes \boldsymbol{v}_m^*),
 \end{equation}
where $\otimes$ denotes the outer product. The strength and sign are controlled by the input parameters $f_m = -\hbar \Delta_{c,m} \eta_m^2/(\Delta_{c,m}^2+\kappa_m^2)$. Since these parameters depend on detuning and amplitude of the pump lasers, one can externally manipulate $A$ without any change of the setup.

So far we have a quite general coupled quantum oscillator implementation in which the state of each oscillator is given by the occupation number at a lattice site. By increasing the on-site repulsion the oscillators get nonlinear and the extra energy required for multiple occupation of a site becomes large. Consequently, for low enough densities only zero or single occupations occur and the bosonic creation and annihilation operators can be mapped to spin-1/2 operators, identifying an occupied site with spin-up and an empty site with spin-down. In this so-called Tonks-Girardeau limit ($U \gg J,\zeta$) the system reduces to a coupled spin model
\begin{equation}
\label{eq:spinhamiltonian}
\begin{aligned}
H_\mathrm{sp} = & - J \sum_i (\sigma_{i+1}^\dagger \sigma_i + \sigma_i^\dagger \sigma_{i+1})\\
	     &- \frac{\zeta}{4} \left( \sum_{i,j} A_{ij} \sigma_i^z \sigma_j^z + \sum_{i} \left[ 2 \sum_j A_{ij} \right] \sigma_i^z \right),
\end{aligned}
\end{equation}
which amounts to the substitutions $b_i \equiv \sigma_i$ and consequently $\hat n_i \equiv \frac{1}{2} ( \sigma_i^z + 1 )$, where $\sigma_i^\alpha$ are Pauli matrices and $\sigma_i = \frac{1}{2} (\sigma_i^x-i\sigma_i^y)$. Formally it is a projection of $H_\mathrm{ad}$ onto the zero and single occupation subspace, which is valid within first order perturbation theory in the small parameters $J/U$ and $\zeta/U$ \cite{mila2011strong}. Note that this limit is already reached for moderate lattice depths $V_\mathrm{L} \approx 10 E_\mathrm{R}$ \cite{paredes2004tonks}. An equivalent model appears for polarized fermions in the lattice.

Since $H_\mathrm{sp}$ commutes with $\sum_i \sigma_i^z$ the accessible Hilbert space reduces to the $\binom{N}{N_\mathrm{A}}$-dimensional subspace with fixed number of spin-up particles. However, for $N_\mathrm{A} = N/2$ the subspace still grows exponentially with $N$.

\section{Constructing an interaction matrix}
\label{sec:interactionmatrix}
Let us now investigate how to realize a general interaction matrix $A$. While its off-diagonal elements determine the interaction between two pseudo-spins, the diagonal elements specify the local field strengths in the second line of Equation (\ref{eq:spinhamiltonian}). Specifically, a local field strength on the $i$-th spin $h_i = 2 \sum_j A_{ij}$ corresponds to the diagonal element $A_{ii} = h_i/2 - \sum_{j \neq i} A_{ij}$ in the matrix. Hence, in order to have full control over interactions and local fields we have to specify up to $N(N+1)/2$ elements, which in the worst case requires as many lasers. Fortunately, these are classical fields with fixed amplitude and frequency.

Formally, the interaction matrix (\ref{eq:interactionmatrix}) appears as linear combination of matrices $V_m$ with coefficients $f_m/\zeta$. Thus if we manage to choose mode functions $u_{c,m}$, pump fields $u_{p,m}$ and lattice location such that $\{V_m\}_{m=1,...,N(N+1)/2}$ forms a basis of the real symmetric matrices, Equation (\ref{eq:interactionmatrix}) can be inverted to fix the required input parameters
\begin{equation}
\label{eq:coefficients}
f_m(A) = \zeta \sum_n (G^{-1})_{mn} \langle V_n, A \rangle.
\end{equation}
Here $G$ is the Gram matrix $G_{mn} = \langle V_m, V_n \rangle$ with inner product $\langle A, B \rangle = \mathrm{Tr}(AB^\dagger)$.
In other words, once a set of modes forming a basis is found, we can directly determine the pump laser properties to realize an arbitrary interaction matrix $A$. While $N(N+1)/2$ lasers are needed to get a complete basis set, many interesting interaction matrices can be constructed with a lot less modes.

\section{Quantum Annealing}
\label{sec:quantumannealing}
In principle our setup realizes an effective spin Hamiltonian with general time dependent all-to-all spin interactions and local fields. This allows for quantum simulation and encoding classical optimization problems in its ground state. The numerically non-trivial task of finding the ground state of a Hamiltonian $H_\mathrm{pr}$ is tackled by quantum annealing \cite{kadowaki1998quantum,farhi2001quantum,santoro2006optimization}, which might promise a speedup over classical methods \cite{boixo2014evidence,heim2015quantum,katzgraber2015seeking}. To this end one adiabatically evolves the system with a time-dependent Hamiltonian
\begin{equation}
H_\mathrm{QA}(t) = a(t) H_\mathrm{kin} + b(t) H_\mathrm{pr}.
\end{equation}
The kinetic term $H_\mathrm{kin}$ is chosen simple enough to posses a known gapped ground state. Initially at $t=0$, the first term is dominant, i.e.\ $a(0) \gg b(0)$ and the system is prepared in this ground state of $H_\mathrm{kin}$. By slowly decreasing $a(t)$ and increasing $b(t)$ the second term becomes dominant after an annealing time $\tau$, i.e.\ $a(\tau) \ll b(\tau)$. Due to the adiabatic theorem \cite{kato1950adiabatic} the system approximately stays in its instantaneous eigenstate and thus finally ends up in the ground state of $H_\mathrm{pr}$, provided $\tau$ is large enough, i.e.\ the adiabatic passage is slow. 

The Hamiltonian $H_\mathrm{sp}$ given in Equation (\ref{eq:spinhamiltonian}) with time-dependent coefficients $\zeta(t)$ and $J(t)$ already has the genuine form of a quantum annealing Hamiltonian $H_\mathrm{QA}$, where the first line corresponds to $H_\mathrm{kin}$ and the second line to $H_\mathrm{pr}$. For an adiabatic transfer we ramp up $\zeta(t)$ from $\zeta(0) = 0$ until the kinetic term becomes negligible $\zeta(\tau) \gg J$. This can be achieved by uniformly increasing all $|f_m|$'s, which physically amounts to (i) increasing the strengths of all pump lasers or (ii) tuning them closer to resonance with the cavity modes. The uniformity guarantees that $A$ and thus the structure of $H_\mathrm{pr}$ is not changed during the sweep. A simultaneous increase of the lattice depth to reduce tunneling $J$ helps further.

Note that instead of adiabatic transfer one could implement cavity cooling for the full interacting Hamiltonian to cool towards the ground state starting from a thermal state. This has proven successful for the single mode case \cite{sandner2015self,wolke2012cavity} and cooling profits from more modes \cite{ritsch2013cold}.

\subsection*{Readout}

The final state readout can be done by analyzing the light leaking out from the cavity \cite{mekhov2009quantum,mekhov2012quantum}, where the quantities of interest are the (classical) spins $\langle \sigma_i^z \rangle \equiv 2 \langle \hat n_i \rangle - 1$, which can be calculated from the occupations $\langle \hat n_i \rangle$.

Measuring the output fields $\propto \langle a_m \rangle$ (e.g.\ by homodyne detection) one has to approximately solve the expectation value version of Equations (\ref{eq:fieldeq}) for $\langle \hat n_i \rangle$, which is an overdetermined $M \times N$ linear system of equations, e.g.\ by using a least mean square method. Alternatively, by measuring the output intensities $\propto \langle a_m^\dagger a_m \rangle$ one has to invert
\begin{equation}
\langle a_m^\dagger a_m \rangle = \frac{\eta_m^2}{\Delta_{c,m}^2+\kappa_m^2} \sum_{i,j} V_m^{ij} \langle \hat n_i \hat n_j \rangle
\end{equation}
to obtain the $N(N+1)/2$ correlations $\langle \hat n_i \hat n_j \rangle$. Since in the large-$U$ limit it holds that $\hat n_i^2 \equiv \hat n_i \equiv (\sigma_i^z+1)/2$, the occupations correspond to the diagonal elements $\langle \hat n_i^2 \rangle$.

\section{Associative memory}
\label{sec:associativememory}
As a generic example we consider a Hopfield associative memory network with a quantum annealing recall \cite{santra2016exponential,hopfield1982neural,hopfield1984neurons}. A Hopfield net consists of $N$ binary state units (so-called neurons), which can be represented by (classical) Ising spins $s_i$ interconnected by real symmetric weights $W_{ij}$. For their dynamics Hopfield proposed an iterative update rule, which locally minimizes an energy function $E(\boldsymbol{s}) = - \sum_{i<j} W_{ij} s_i s_j$ of the system state vector $\boldsymbol s = (s_1,...,s_N)$. In combination with a learning rule determining the weights  $W_{ij}$ the network works as an associative memory, which can memorize a set of $P$ states $\mathcal{M} = \{ \boldsymbol{w}_p \}_{p=1,...,P}$. That is, the system converges to the stored state in $P$ having maximal overlap with an initial (input) state. A proven standard choice of weights is provided by the Hebbian learning rule \cite{hebb2005organization}
\begin{equation}
W_{ij} = \frac{1}{P}\sum_{p=1}^P w_p^i w_p^j.
\end{equation}

Each associative memory of size $N$ has a limited capacity, i.e.\ a maximal number of stored states which can be reliably recalled. This capacity grows proportional to $N$ using the aforementioned update rule \cite{amit1987statistical}. Thus convergence to a particular memory state is not guaranteed to succeed for an input state with too strong deviations or if too many states are stored.

This capacity is suggested to scale much more favorable in a quantum simulator version of the model \cite{santra2016exponential}. In such a setup one replaces Hopfield's classical spin update dynamics by quantum annealing to find the ground state of the Hamiltonian 
\begin{equation}
\label{eq:assmemory}
H_\mathrm{AM} = - \sum_{i<j} W_{ij} \sigma_i^z \sigma_j^z - \nu \sum_i \chi_i \sigma_i^z.
\end{equation}

A state of the network $\boldsymbol s$ now corresponds to eigenstates of the $\sigma_i^z$-operators $|\boldsymbol s \rangle$. Obviously, the first term is the pendant to the energy function $E(\boldsymbol s)$, which lowers the energy of memory states $| \boldsymbol w_p \rangle$. The input state $\boldsymbol \chi$ is encoded in the local fields (as opposed to the classical case, where it is the initial state), such that the energy of a state $|\boldsymbol s \rangle$ is lowered proportionally to its similarity to $| \boldsymbol \chi \rangle$ quantified by the inner product $\boldsymbol \chi \cdot \boldsymbol s = \sum_i \chi_i s_i$. The ground state then corresponds to the memorized state with maximal overlap with $\boldsymbol \chi$ for a not too large $\nu$ as discussed in \cite{santra2016exponential} (see also Appendix \ref{app:associativememory}).

In our system $H_\mathrm{AM}$ can be realized with the interaction matrix
\begin{equation}
\label{eq:hopfieldintmatrix}
A_{ij} = W_{ij} + \nu \chi_i \delta_{ij}
\end{equation}
in the coupled spin Hamiltonian of Equation (\ref{eq:spinhamiltonian}), where $\delta_{ij}$ denotes the Kronecker delta. Physically each lattice site corresponds to a neuron with the two states `occupied' and `not occupied' and weights are determined by the pump lasers and cavity modes.

\begin{figure}
\includegraphics[scale=1]{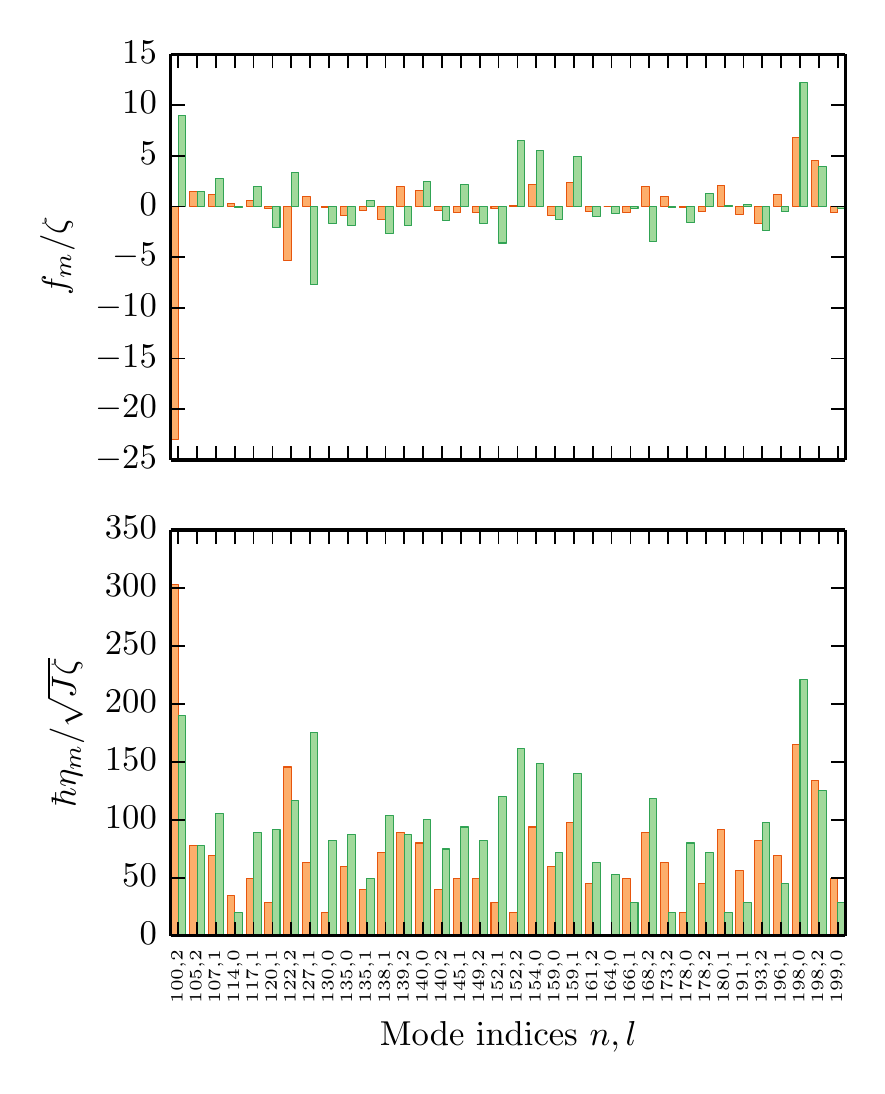}
\caption{\textbf{Top:} Input parameters for the two input states $\boldsymbol{\chi}_1$ (orange) and $\boldsymbol{\chi}_2$ (green) given in the main text for a chosen set of modes $\mathcal{B}$, where the second transverse mode index $m = 0$. \textbf{Bottom:} These input parameters can be realized by the pump strengths $\eta_m = \sqrt{-f_m(\Delta_{c,m}^2+\kappa_m^2)/(\hbar \Delta_{c,m})}$, where $\kappa_m = 1000 J / \hbar$ and $\Delta_{c,m} = \mathrm{sgn}(-f_m) \kappa_m$. Due to weak coupling e.g.\ the first mode needs to be pumped strong in both cases.}
\label{fig:fm}
\end{figure}

\subsection*{Example}

Let us now consider a specific problem with 8 sites ($N=8$) filled with 4 particles resulting in $\binom{8}{4} = 70$ possible states. We want to store two memory states
\begin{equation*}
\begin{aligned}
\boldsymbol{w}_1 &= (1,1,-1,-1,1,-1,1,-1)\\
\boldsymbol{w}_2 &= (1,1,-1,1,1,-1,-1,-1).
\end{aligned}
\end{equation*}
Recalling the input patterns 
\begin{equation*}
\begin{aligned}
\boldsymbol{\chi}_1 &= (1,1,1,-1,-1,-1,1,-1)\\
\boldsymbol{\chi}_2 &= (1,1,-1,1,-1,-1,-1,1)
\end{aligned}
\end{equation*}
and choosing $\nu = 0.7$ amounts to specifying the interaction matrices $A_{\chi_1}$ and $A_{\chi_2}$ (see Appendix \ref{app:interactionmatrices}). The similarities between the states are summarized by $\boldsymbol \chi_i \cdot \boldsymbol w_j = 4 \delta_{ij}$. Thus we can already anticipate the expected results: Upon recalling $\boldsymbol \chi_1$ ($\boldsymbol \chi_2$) the ground state of the system should converge to $\boldsymbol w_1$ ($\boldsymbol w_2$) for large $\zeta/J$.

In the following we go through the steps for implementing such a problem in our system: Firstly, we search for a `good' choice of modes and geometry for this system size. Secondly, we implement the stated problem, i.e.\ the interaction matrices for $\boldsymbol \chi_1$ and $\boldsymbol \chi_2$. Finally, we simulate the coherent annealing dynamics which should yield the solution to the problem.

\begin{figure}
\includegraphics[scale=1]{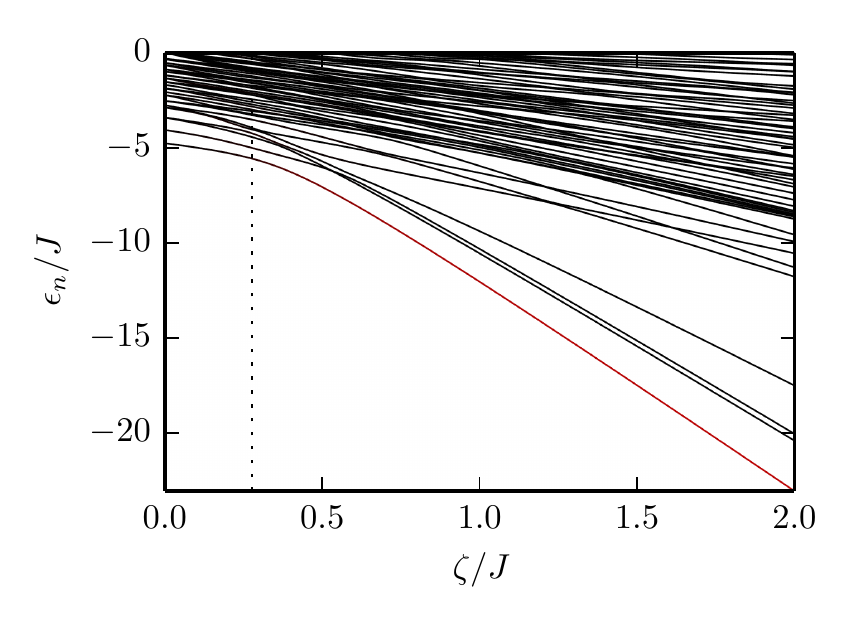}
\caption{Spectrum of the lowest few eigenvalues $\epsilon_n$ of $H_\mathrm{sp}$, Equation (\ref{eq:spinhamiltonian}), as function of $\zeta$ for a recall of the input pattern $\boldsymbol{\chi}_1$, leading to the recovered interaction matrix $\tilde A_{\chi_1}$. The dotted line at $\zeta/J = 0.28$ shows the position of the smallest gap between ground state and first excited state, while the color of the lines encodes the overlap of the target memory state with the eigenstates $|\langle \phi_n(\zeta) | \boldsymbol w_1 \rangle |^2$ from black ($=0$) to red ($=1$). Already at $\zeta/J = 2$, the ground state is very close to the target state: $|\langle \phi_0(\zeta=2J) | \boldsymbol w_1\rangle |^2 = 0.976$.}
\label{fig:spectrum1}
\end{figure}

\textit{A specific set-up.}---We consider a cavity supporting several Hermite-Gaussian modes denoted by the longitudinal mode index $n$ and the transverse mode indices $l$ and $m$, which define the transverse cavity axes $r_l$ and $r_m$. The external 1D optical lattice has a depth of $V_\mathrm{L} = 10 E_\mathrm{R}$ and a spacing $d = 1.2 \lambda_{n=100}/2$ and is located in the $z$-$r_l$-plane of the cavity (see Figure \ref{fig:setup}). The standing wave pump lasers are approximated by plane waves and are applied orthogonally to the lattice axis such that each has an anti-node at the lattice location and consequently $u_{p,m}(x) = 1$. Thus the form of the couplings between lattice and modes only depends on which cavity modes (indexed by $n,l,m$) are addressed and where the lattice is positioned, i.e.\ where the cavity mode functions are evaluated. The ratio of radius of curvature of the mirrors and cavity length is chosen as $R/L = 2/3$.

\textit{Finding the best suitable modes.}---In order to invert Equation (\ref{eq:interactionmatrix}) one has to choose $N(N+1)/2 = 36$ linearly independent single-mode coupling matrices $V_m$, i.e.\ $\langle V_m, V_n \rangle \neq 0$, forming a basis $\mathcal{B}$ of the matrix space. Due to the different spatial shape of the mode functions this is generally fulfilled for most mode choices in principle. However, if the $V_m$'s are too similar, in practise an unrealistically high precision for the input laser parameters $f_m$ is needed to reliably implement the most general interaction matrix. Therefore, to reduce the experimental restrictions on laser control, one should find a set of modes, which gives rise to a distinct set of single-mode coupling matrices. As a figure of merit one can use the determinant of the Gram matrix of the normalized $V_m$'s (i.e.\ the squared volume spanned by those vectors), which should be maximized (orthogonal vectors would lead to the maximal value of 1). Additionally, we optimize over different lattice orientations (for more details see Appendix \ref{app:modeselection}). Here we restrict ourselves to modes from the candidate set $n \in \{100,199\}$, $l \in \{0,1,2\}$ and $m = 0$.

Let us emphasize that this step is only needed due to the finite precision available and crucially depends on the specific implementation. The modes do not have to be optimal, but only sufficiently `good' for the given precision of the input parameters.

\begin{figure}
\includegraphics[scale=1]{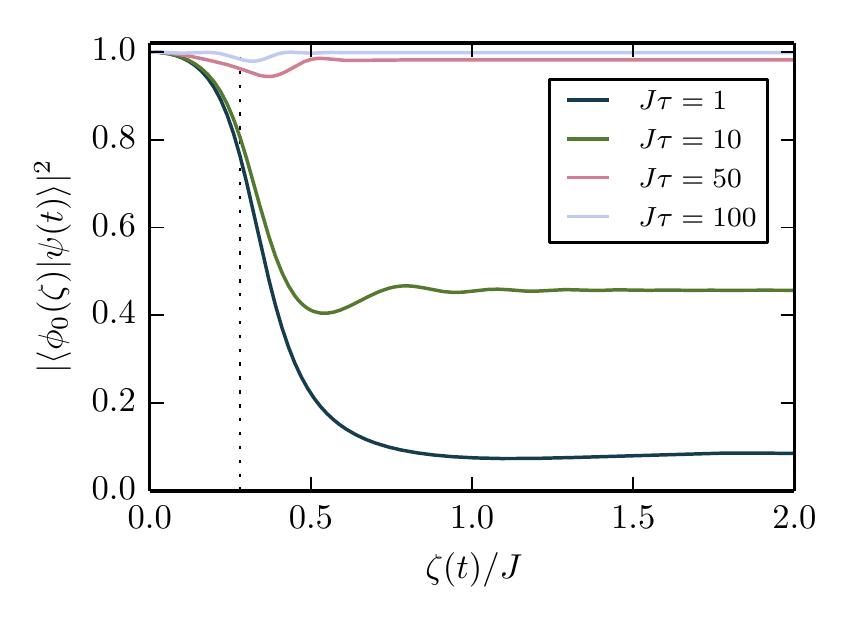} \caption{Time evolution of the overlap between the instantaneous ground state $| \phi_0(\zeta) \rangle$ and the solution of the time-dependent Schr\"odinger equation $|\psi(t)\rangle$ for the linear ramp $\zeta(t) = 2 J t/\tau$ and different annealing times $\tau$. The vertical dotted line shows the location of the smallest gap, as in Figure \ref{fig:spectrum1}.}
\label{fig:gsov1}
\end{figure}

\textit{Input parameters.}---Choosing the modes (and thus fixing $\mathcal B$) has to be done only once for a certain system size $N$. Afterwards any specific interaction matrix can be realized by changing the input parameters $f_m(A)$. We calculate these parameters from Equation (\ref{eq:coefficients}) for $A_{\chi_1}$ and $A_{\chi_2}$ and subsequently round to one decimal place yielding $\tilde f_m^{\chi_i}$, mimicking some finite maximally possible experimental accuracy. The recovered interaction matrix $\tilde A_{\chi_i} = A(\tilde f_1^{\chi_i},...,\tilde f_M^{\chi_i})$ will then approximate $A_{\chi_i}$ depending on how well we chose the modes and how accurately we impose the input parameters. The upper plot in Figure \ref{fig:fm} shows the input parameters for $\tilde A_{\chi_1}$ and $\tilde A_{\chi_2}$, which can be realized by the pump strengths $\eta_m$ shown in the lower plot, assuming the same $|\Delta_{c,m}|$ for each mode.

\begin{figure}
\includegraphics[scale=1]{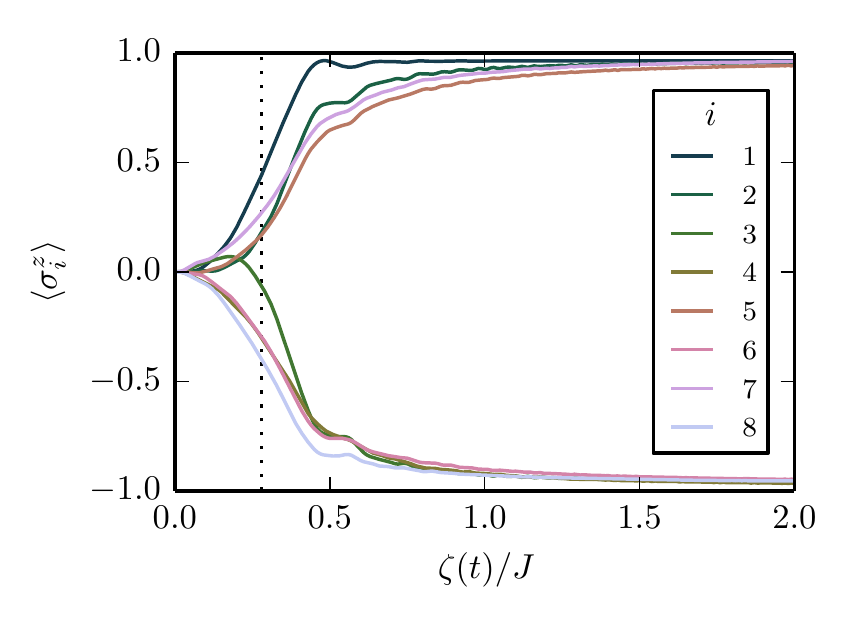}
\caption{Time evolution of the expectation values $\langle \sigma_i^z \rangle$ for each lattice site $i$ for an annealing time of $J \tau = 50$ and a linear ramp $\zeta \propto t$ (see Figure \ref{fig:gsov1}). The overlap with the target state in the end is $|\langle \psi(\tau) | \boldsymbol w_1 \rangle|^2 = 0.959$. Due to the finite annealing time there is a fraction in the excited states and thus the curves do not converge to 1 and -1 exactly.}
\label{fig:occ}
\end{figure}

\textit{Adiabatic passage.}---These approximate interaction matrices define the Hamiltonian $H_\mathrm{sp}(\zeta)$ as a function of $\zeta$, whose eigenvalue spectrum is shown in Figure \ref{fig:spectrum1} for $\boldsymbol \chi_1$. With increasing $\zeta / J$ the ground state converges to $|\boldsymbol w_1 \rangle$ since the inner product $\boldsymbol \chi_1 \cdot \boldsymbol w_1 = 4$ is larger than $\boldsymbol \chi_1 \cdot \boldsymbol w_2 = 0$. This can be already seen at $\zeta/J = 2$, where the overlap between ground state and target state is $|\langle \phi_0(\zeta=2J) | \boldsymbol w_1\rangle |^2 = 0.976$. We observe that the minimum gap between ground and first excited state is $\delta_\mathrm{min} = 0.56 J$ at $\zeta = 0.28 J$. During a time evolution with increasing $\zeta$ this is the most likely region for Landau-Zener tunneling from ground state to excited states.

The typical behavior of the time-dependent solution of the Schr\"odinger equation for a linear sweep and different annealing times $\tau$ is shown in Figure \ref{fig:gsov1}, where we see that for $J\tau \gtrsim 50$ the system stays close to the ground state in this specific example. Especially, the final overlap with the target state $\boldsymbol w_1$ for $J\tau=50$ is $|\langle \psi(\tau) | \boldsymbol w_1 \rangle|^2 = 0.959$. This can also be seen from the time evolution of the individual spins $\langle \sigma_i^z \rangle$ as depicted in Figure \ref{fig:occ}: From an initially unpolarized configuration, they evolve to a value close to $1$ or $-1$ corresponding to $\boldsymbol w_1$. The annealing time $J\tau=50$ translates to $\tau = 100 \, \mathrm{ms}$ for $^{87}$Rb with $E_\mathrm{R} / \hbar \approx 24 \, \mathrm{kHz}$ and $J \approx 0.02 E_\mathrm{R}$, which is a realistic ramp time \cite{landig2016quantum}.

\begin{figure}
\includegraphics[scale=1]{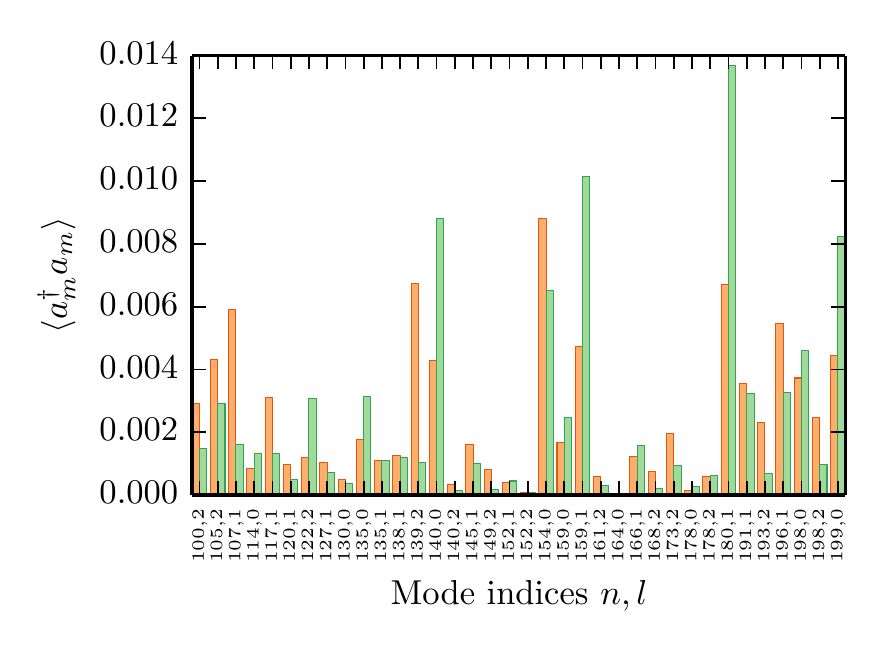}
\caption{Different atomic states give rise to distinct intensity patterns, which can be measured. Here they are shown for the states $\boldsymbol w_1$ (orange) and $\boldsymbol w_2$ (green) at $\zeta/J = 2$. The parameters are as given in Figure \ref{fig:fm}.}
\label{fig:int}
\end{figure}

At the end of the ramp when we have prepared the final state, it can be directly determined in a non-destructive way by measuring the output intensities shown in Figure \ref{fig:int}. This is a crucial advantage of our open system architecture compared close atomic lattice implementation, where site resolved atomic detection is required at the end.  


\section{Conclusions}

We demonstrated how to obtain a coupled Ising spin model from a dilutely filled optical lattice within a multimode cavity with the help of transverse pump lasers. The interactions and local fields of the spins can be tuned by changing the power and detuning of the lasers allowing for real time control. This can be used to slowly ramp up the spin-spin interactions, implementing a quantum annealing dynamics. The final atomic state can be nearly non-destructively read out by measuring the cavity output fields.

Let us point out that the system studied here is technologically not far from current available experimental configurations as used at ETH \cite{landig2016quantum} and Hamburg \cite{kessler2014steering}. These need to be extended by adding extra laser frequencies, as provided by existing frequency comb and amplifier technology. As cavity and comb modes are equidistant, a single lock would be sufficient to bring all modes to resonance. While the general quadratic scaling of the number of lasers with the lattice sites number seems to be rather restrictive at first, the lasers are just a classical resource here. It also turns out that the required number of laser frequencies for a specific problem can be strongly reduced by applying the same laser from different angles.

In our example we found the desired state via adiabatic transfer. As said, for our open system, adiabatic transfer is not the only possibility as the ground state can also be reached via cavity side band cooling \cite{sandner2015self,wolke2012cavity} generalized to the multimode case. In this case the scan time can be reduced as errors are corrected by cooling at a later stage. 

{\bf Acknowledgements.} We thank W.\ Lechner, T.\ Donner, J.\ Leonard and G.\ De las Cuevas for helpful discussions. This work is supported by the Austrian Science Fund Project I1697-N27.

\appendix

\section{Interaction matrices}
\label{app:interactionmatrices}

Recalling the pattern $\boldsymbol{\chi}_1 = (1,1,1,-1,-1,-1,1,-1)$ and choosing $\nu = 0.7$ results in an interaction matrix $A_{\chi 1}$ given by 
\begin{equation*}
\left(
\begin{matrix}

1.7 & 1.0 & -1.0 & 0.0 & 1.0 & -1.0 & 0.0 & -1.0 \\
1.0 & 1.7 & -1.0 & 0.0 & 1.0 & -1.0 & 0.0 & -1.0 \\
-1.0 & -1.0 & 1.7 & 0.0 & -1.0 & 1.0 & 0.0 & 1.0 \\
0.0 & 0.0 & 0.0 & 0.3 & 0.0 & 0.0 & -1.0 & 0.0 \\
1.0 & 1.0 & -1.0 & 0.0 & 0.3 & -1.0 & 0.0 & -1.0 \\
-1.0 & -1.0 & 1.0 & 0.0 & -1.0 & 0.3 & 0.0 & 1.0 \\
0.0 & 0.0 & 0.0 & -1.0 & 0.0 & 0.0 & 1.7 & 0.0 \\
-1.0 & -1.0 & 1.0 & 0.0 & -1.0 & 1.0 & 0.0 & 0.3

\end{matrix}
\right).
\end{equation*}

Using the above modes this matrix can be realized by the following laser input parameters
\begin{equation*}
\begin{aligned}
\boldsymbol{\tilde f}^{\chi_1}/\zeta = (&-23.,    1.5,   1.2,   0.3,   0.6,  -0.2,  -5.3,   1.,   -0.1, \\
			 &-0.9,  -0.4,  -1.3, 2.,    1.6,  -0.4,  -0.6,  -0.6,\\
			 &  -0.2,   0.1,   2.2,  -0.9,   2.4,  -0.5,  0., -0.6,   2.,   1., \\
			 &   -0.1,  -0.5,   2.1,  -0.8,  -1.7,   1.2,   6.8,   4.5,  -0.6),
\end{aligned}
\end{equation*}
which are already rounded to one position after the decimal point. We see that all parameters have similar magnitude, which is due to the proper choice of the modes. The recovered interaction matrix from the rounded input parameters $\tilde A_{\chi_1}$ is (rounded up to 2 positions after decimal point)

\begin{equation*}
\left(
\begin{matrix}

1.72 & 1.01 & -1.01 & 0.02 & 0.99 & -1.00 & 0.00 & -0.97 \\
1.01 & 1.67 & -0.91 & -0.02 & 0.99 & -0.99 & 0.00 & -1.01 \\
-1.01 & -0.91 & 1.66 & 0.03 & -1.00 & 0.98 & 0.01 & 1.01 \\
0.02 & -0.02 & 0.03 & 0.27 & 0.05 & 0.03 & -1.00 & -0.00 \\
0.99 & 0.99 & -1.00 & 0.05 & 0.33 & -0.97 & -0.03 & -1.00 \\
-1.00 & -0.99 & 0.98 & 0.03 & -0.97 & 0.29 & 0.01 & 0.96 \\
0.00 & 0.00 & 0.01 & -1.00 & -0.03 & 0.01 & 1.70 & 0.02 \\
-0.97 & -1.01 & 1.01 & -0.00 & -1.00 & 0.96 & 0.02 & 0.30

\end{matrix}
\right)
\end{equation*}
which is similar to $A_{\chi_1}$.

Recalling another pattern $\boldsymbol{\chi}_2 = (1,1,-1,1,-1,-1,-1,1)$ results in an interaction matrix which differs from $A_{\chi_1}$ only in the diagonal (since the memory is the same), i.e. $A_{\chi_2} = $

\begin{equation*}
\left(
\begin{matrix}

1.7 & 1.0 & -1.0 & 0.0 & 1.0 & -1.0 & 0.0 & -1.0 \\
1.0 & 1.7 & -1.0 & 0.0 & 1.0 & -1.0 & 0.0 & -1.0 \\
-1.0 & -1.0 & 0.3 & 0.0 & -1.0 & 1.0 & 0.0 & 1.0 \\
0.0 & 0.0 & 0.0 & 1.7 & 0.0 & 0.0 & -1.0 & 0.0 \\
1.0 & 1.0 & -1.0 & 0.0 & 0.3 & -1.0 & 0.0 & -1.0 \\
-1.0 & -1.0 & 1.0 & 0.0 & -1.0 & 0.3 & 0.0 & 1.0 \\
0.0 & 0.0 & 0.0 & -1.0 & 0.0 & 0.0 & 0.3 & 0.0 \\
-1.0 & -1.0 & 1.0 & 0.0 & -1.0 & 1.0 & 0.0 & 1.7

\end{matrix}
\right).
\end{equation*}

Analogously, it can be implemented by the rounded input parameters
\begin{equation*}
\begin{aligned}
\boldsymbol{\tilde f}^{\chi_2}/\zeta = (&  9.,    1.5,   2.8 , -0.1,   2.,   -2.1 ,  3.4,  -7.7,  -1.7, -1.9,\\
								&   0.6,  -2.7,  -1.9,   2.5,  -1.4,   2.2,  -1.7, -3.6,   6.5,  \\
							  & 5.5,  -1.3,   4.9,  -1.,   -0.7,  -0.2,  -3.5, -0.1, \\
							  & -1.6,   1.3,   0.1,   0.2,  -2.4,  -0.5,  12.2,   3.9,  -0.2).
\end{aligned}
\end{equation*}

\section{Selecting the modes}
\label{app:modeselection}

\begin{figure}
\includegraphics[width=1\columnwidth]{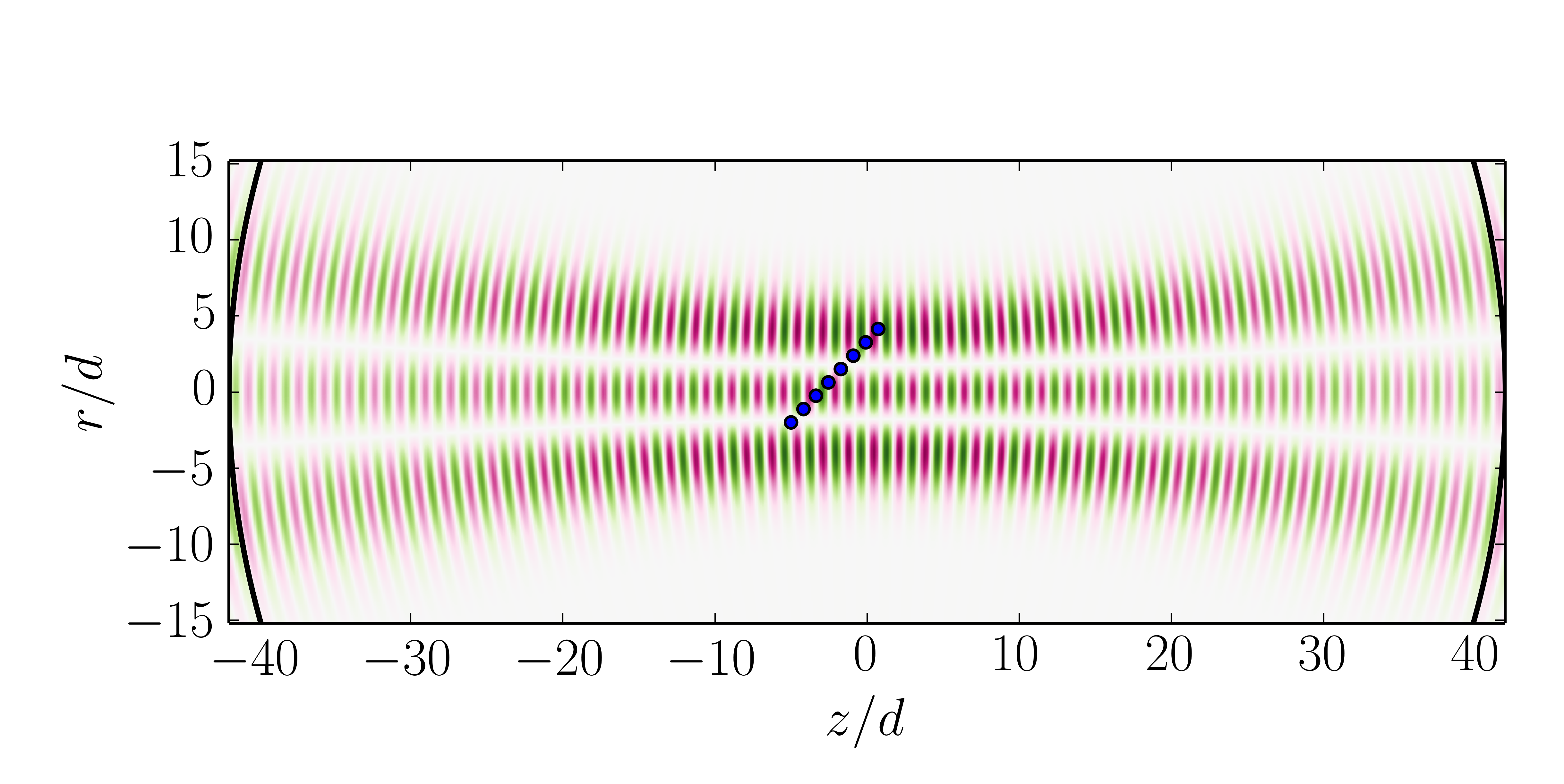}
\caption{The lattice location within the cavity used in the example in Section \ref{sec:associativememory}. Blue dots indicate positions of individual lattice sites. The depicted cavity mode is $(n,l)=(100,2)$. The black arcs are the cavity mirrors.}
\label{fig:mode}
\end{figure}

As discussed in Section \ref{sec:associativememory}, in the set-up of this specific example the single-mode coupling matrices $V_m$ depend on the cavity modes only. Thus in order to find good modes we choose a candidate set of $K=300$ Hermite-Gauss cavity modes with longitudinal mode indices $n \in \{100,199\}$ and transverse mode indices $l \in \{0,1,2\}$ and $m=0$. Now we aim to find a subset of $M = N(N+1)/2 = 36$ modes which results in a large determinant of the Gram matrix created by the normalized $V_m$'s for one specific lattice location (it does not have to be the optimum). Since the number of $M$-combinations out of the candidate set of size $K$ is huge, $\binom{K}{M} \sim 10^{46}$, we cannot try out all, but have to use some algorithm which still scales polynomially. We use one possible choice which is given by:
\begin{itemize}
\item[(i)] Compare all pairs of modes and choose the best ($K(K-1)$ steps).
\item[(ii)] Subsequently add the best mode until ending up with $M$ modes ($< MK$ steps).
\item[(iii)] Tentatively replace each selected mode by one mode of the remaining candidate set and take the best replacement, but only if the new Gram matrix determinant is larger. Repeat this for all selected modes ($M(K-M)$ steps).
\end{itemize}
In each step, the Gram matrix determinant has to be calculated. We repeat this procedure for different lattice locations and angles to the cavity axes. In addition to that, we post-select sets of modes where the norms $\sqrt{\langle V_m, V_m \rangle}$ are relatively uniform in order to guarantee uniform input parameters.

A resulting `good' lattice location is given by the coordinates of the first lattice site $z^0 = -5 d$, $r_l^0 = -2d$ and $r_m^0 = 0$ in a coordinate system with origin at the cavity center and has an angle $\phi = 47^\circ$ with respect to the cavity $z$-axis (see Figure \ref{fig:mode}). The set of selected modes for this lattice location, defining the basis $\mathcal{B}$, is given as mode index tuples $(n,l)$
\begin{equation*}
\begin{aligned}
\mathcal{B} = \\
\{&(100, 2), (105, 2), (107, 1), (114, 0), (117, 1), (120, 1),\\
&(122, 2), (127, 1), (130, 0), (135, 0), (135, 1), (138, 1),\\
&(139, 2), (140, 0), (140, 2), (145, 1), (149, 2), (152, 1),\\
&(152, 2), (154, 0), (159, 0), (159, 1), (161, 2), (164, 0),\\
&(166, 1), (168, 2), (173, 2), (178, 0), (178, 2), (180, 1),\\
&(191, 1), (193, 2), (196, 1), (198, 0), (198, 2), (199, 0) \}.
\end{aligned}
\end{equation*}
This configuration results in a Gram matrix determinant of $3.21 \times 10^{-11}$.

\section{Adiabatic elimination of the cavity modes}
\label{app:adiabaticelimination}
The adiabatic elimination of the cavity modes has already been discussed in similar set-ups, see e.g.\ \cite{nagy2009nonlinear,maschler2008ultracold,habibian2013bose}. The Heisenberg-Langevin equation of the cavity field operators is
\begin{equation}
\begin{aligned}
\dot a_m &= \frac{1}{i \hbar} [a_m, H] \\
								  &= (i \delta_{c,m} - \kappa_m) a_m - i \eta_m \sum_i v_m^i \hat n_i + \hat \xi_m.
\end{aligned}
\end{equation}
Including the coupling of the cavity modes with the vacuum field gives rise to field decay and an input noise operator $\hat \xi_m$ with $\langle \hat \xi(t) \rangle = 0$ and $\langle \hat \xi(t), \hat \xi(t') \rangle = 2\kappa_m \delta(t-t')$.
Formal integration from $t_0$ to $t$ leads to
\begin{equation}
\label{eq:formalintegration}
\begin{aligned}
a_m(t) =& e^{(i \delta_{c,m}-\kappa_m) \Delta t} a_m(t_0) \\
		&- i \eta_m \sum_i v_m^i \int_0^{\Delta t} \mathrm ds \, e^{(i\delta_{c,m}-\kappa_m)s} \hat n_i (t-s)\\
		&+ \hat \Sigma_m(t)
\end{aligned}
\end{equation}
with the new noise operator
\begin{equation}
\hat\Sigma_m(t) = \int_0^{\Delta t} \mathrm ds \, e^{(i\delta_{c,m}-\kappa_m)s} \hat \xi_m(t-s).
\end{equation}
The time step $\Delta t = t-t_0$ defines an intermediate time scale: One the one hand it is (i) much larger than the cavity time scale $\kappa_m^{-1} \ll \Delta t$, and on the other hand (ii) much smaller than the time scale of the atomic motion $\Delta t \ll (J/\hbar)^{-1}$.

Due to (i) we can neglect the first term in (\ref{eq:formalintegration}). Moreover, because of (ii) the atomic operator $\hat n_i$ does not vary much in the time $\Delta t$ and can hence be approximated by $\hat n_i(t)$, which allows us to evaluate the integral. This approximation amounts to truncating an expansion in the small parameter $\dot{\hat{n}}_i/|\delta_{c,m}+i\kappa_m| \propto J / (\hbar|\delta_{c,m}+i\kappa_m|)$ at zeroth order. It yields
\begin{equation}
\label{eq:adfield}
a_m(t) = \frac{\eta_m}{\delta_{c,m}+i\kappa_m} \sum_i v_m^i \hat n_i(t) + \hat \Sigma_m(t).
\end{equation}
Within the limit (i), the noise operator has the properties $\langle \hat \Sigma_m(t) \rangle = 0$ and $\langle \hat \Sigma_m(t) \hat \Sigma_m^\dagger(t') \rangle = \frac{2\kappa_m}{\delta_{c,m}^2+\kappa_m^2}\delta(t-t')$ \cite{nagy2009nonlinear}.


The interaction part of the Heisenberg equation of a the bosonic annihilation operator is
\begin{equation}
\dot b_i = -i \sum_m \eta_m ( v_m^i a_m^\dagger b_i + (v_m^i)^* b_i a_m),
\end{equation}
where a specific order of atomic and cavity operators was chosen. The ordering freedom leads to ambiguities \cite{maschler2008ultracold}. Plugging in (\ref{eq:adfield}) without the noise term yields
\begin{equation}\label{eq:heisenbergbi}
\begin{aligned}
\dot b_i =& -i \sum_m \frac{\delta_{c,m} \eta_m^2}{\delta_{c,m}^2+\kappa_m^2} (v_m^i(v_m^j)^* \hat n_j b_i + (v_m^i)^* v_m^j b_i \hat n_j) \\
     & + \sum_m \frac{\kappa_m \eta_m^2}{\delta_{c,m}^2+\kappa_m^2} (v_m^i(v_m^j)^* \hat n_j b_i - (v_m^i)^* v_m^j b_i \hat n_j).
\end{aligned}
\end{equation}
Using the identity $[b_i, \sum_{j,k} M_{jk} \hat n_j \hat n_k] = \sum_j (M_{ij} \hat n_j b_i + M_{ji} b_i \hat n_j)$ we realize that the first term can be obtained from $\dot b_i = 1/(i\hbar)[b_i, H_\mathrm{ad}^\mathrm{int}]$ with a purely atomic Hamiltonian
\begin{equation}
\begin{aligned}
H_\mathrm{ad}^\mathrm{int} &= \hbar \sum_{i,j} \sum_m \frac{\delta_{c,m} \eta_m^2}{\delta_{c,m}^2+\kappa_m^2} v_m^i (v_m^j)^* \hat n_i \hat n_j \\
             &= \hbar \sum_{i,j} \sum_m \frac{\delta_{c,m} \eta_m^2}{\delta_{c,m}^2+\kappa_m^2} \mathrm{Re}(v_m^i (v_m^j)^*) \hat n_i \hat n_j,
\end{aligned}
\end{equation}
where we used $[\hat n_i, \hat n_j] = 0$ in the second line.

The incoherent dynamics coming from Lindblad terms $\mathcal{L}\rho = \sum_m (2 C_m \rho C_m - C_m^2 \rho - \rho C_m^2)$ with the Hermitian operators
\begin{equation}
C_m = \sqrt{2 \kappa_m} \frac{\eta_m}{\sqrt{\delta_{c,m}^2+\kappa_m^2}} \sum_i v_m^i \hat n_i
\end{equation}
gives rise to the second term in (\ref{eq:heisenbergbi}) and the noise (which we did not explicitly consider). In the main text we neglect this incoherent contribution, which well describes the physics in current experiments \cite{baumann2010dicke,landig2016quantum}.

\section{Detailed analysis of the associative memory Hamiltonian}
\label{app:associativememory}
We discuss the structure of the Hamiltonian $H_{AM}$, which is described in \cite{santra2016exponential}. Since this Hamiltonian is diagonal in the occupation number basis (it only contains $\sigma^z$-operators), the analysis can be reduced to a classical energy function. The energy of an arbitrary state $\boldsymbol s$ evaluates to
\begin{equation}
\label{eq:energy}
E_\mathrm{AM}(\boldsymbol s) = \langle \boldsymbol s | H_{AM} | \boldsymbol s \rangle = -\frac{1}{2P} \sum_{q=1}^P \langle \boldsymbol s, \boldsymbol w_q \rangle^2 - \nu \langle \boldsymbol s, \boldsymbol \chi \rangle.
\end{equation}
The goal is that the lowest energy state
\begin{itemize}
\item[(i)] is a memory state and
\item[(ii)] has maximum similarity to the input pattern $\boldsymbol \chi$,
\end{itemize}
i.e.\ $\boldsymbol w_k = \max_p \langle \boldsymbol w_p, \boldsymbol \chi \rangle$ with $\boldsymbol w_p \in \mathcal M$. Formally we require
\begin{equation}
\label{eq:groundstatecondition}
E_\mathrm{AM}(\boldsymbol w_k) < E_\mathrm{AM}(\boldsymbol s) \; \mathrm{for} \; \boldsymbol s \neq \boldsymbol w_k := \max_p \langle \boldsymbol w_p, \boldsymbol \chi \rangle.
\end{equation}
While the first term in $E_\mathrm{AM}$ is responsible for requirement (i), the second term should come up for (ii).

\subsection{Memory term}

Let us now consider the first term ($\nu=0$), which lowers the energy of memory states to
\begin{equation}
E_\mathrm{AM}(\boldsymbol w_p) = -\frac{1}{2P} \sum_{q=1}^P \langle \boldsymbol w_p, \boldsymbol w_q \rangle^2 = -\frac{N}{2} - \frac{1}{2P} \sum_{q \neq p} \langle \boldsymbol w_p, \boldsymbol w_q \rangle^2.
\end{equation}
We observe that all memory states are degenerate, i.e.\ $E_\mathrm{AM}(\boldsymbol w_p)$ is independent of $p$, if the dot product of all memory pattern pairs is the same: $\langle \boldsymbol w_p, \boldsymbol w_q \rangle = a$ for all $p \neq q$ and $a \in \mathbb{Z}$. This is guaranteed e.g.\ for pairwise orthogonal memory states $\langle \boldsymbol w_p, \boldsymbol w_q \rangle = N \delta_{pq}$ (i.e.\ $a=0$) and for $P=2$ due to the commutativity of the dot product ($\langle \boldsymbol w_1, \boldsymbol w_2 \rangle = \langle \boldsymbol w_2, \boldsymbol w_1 \rangle$).

\subsection{Recall term}
The second term in (\ref{eq:energy}) lowers the energy of states close to an input pattern $\boldsymbol \chi$. Now we clarify the bounds on the size of this term $\nu$.

\textit{Lower bound.}---If the memory states are degenerate an arbitrarily small $\nu > 0$ is sufficient to bias the memory state with maximum overlap to $\boldsymbol \chi$. In other words, the lower bound on the local field strength is zero, $\nu_\mathrm{min} = 0$, in case of degenerate memory states.

For non-degenerate memories in general we need a lower bound $\nu_\mathrm{min} > 0$ in order to get the right solution, since certain memory patterns will be preferred over others.

\textit{Upper bound.}---Moreover, we have to make sure that the input pattern is not overbiased, i.e.\ that the input pattern itself does not become the ground state in order to meet requirement (i). That is
\begin{equation}
\min_p E_\mathrm{AM}( \boldsymbol w_p ) < E_\mathrm{AM}(\boldsymbol \chi),
\end{equation}
which leads to an upper bound for $\nu$:
\begin{equation}
\nu < \max_p \frac{1}{2P(N-\langle \boldsymbol \chi, \boldsymbol w_p \rangle)} \sum_{q=1}^P \left( \langle \boldsymbol w_p, \boldsymbol w_q \rangle^2 - \langle \boldsymbol \chi, \boldsymbol w_q \rangle^2 \right).
\end{equation}
However, there is a caveat: Calculating this bound amounts to evaluating all inner products $\langle \boldsymbol \chi, \boldsymbol w_q \rangle$, which solves the problem of finding the most similar memory state to $\boldsymbol \chi$ and thus renders the whole annealing procedure superfluous.

\begin{figure}[htp]
\includegraphics{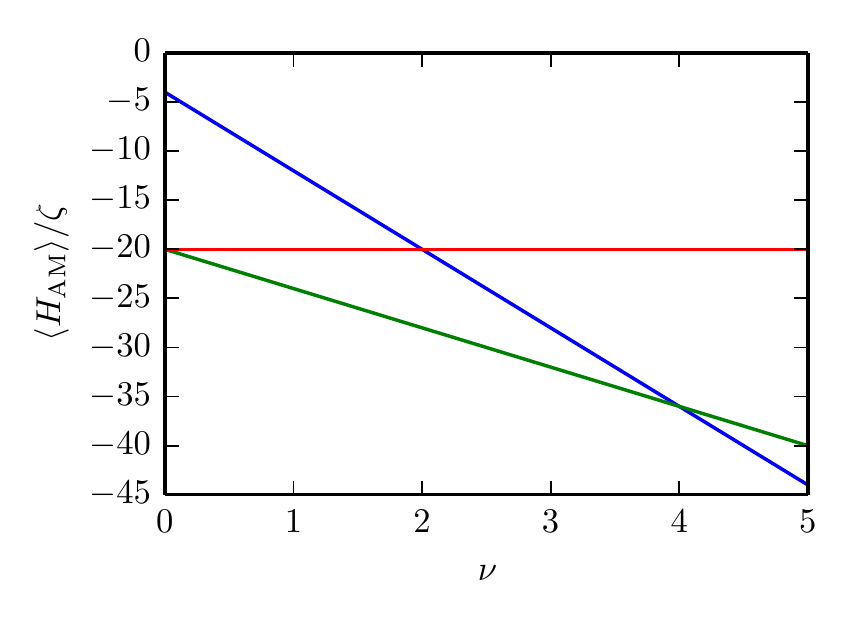}
\caption{The energies depending on the choice of $\nu$ when recalling $\boldsymbol \chi_1$. The recall bias $E_\mathrm{AM}(\boldsymbol \chi_1)$ (blue) has to be smaller than $E_\mathrm{AM}(\boldsymbol w_1)$ (green), hence we need to choose $0 < \nu < 4$. $E_\mathrm{AM}(\boldsymbol w_2)$ (red) is not affected by $\nu$ due to $\langle \boldsymbol \chi_1, \boldsymbol w_2 \rangle = 0$. Here $P=2$ such that the memory patterns are degenerate.}
\label{fig:nu}
\end{figure}

For the special case of degenerate memories however, one can simply choose the smallest possible $\nu > 0$ (depending on the available precision). This situation is depicted in Figure \ref{fig:nu}. Having non-degenerate memories, one could repeat for different values of $\nu$. For large values of $\nu$, the resulting ground state should be $\boldsymbol \chi$. Upon lowering $\nu$ we should arrive at a point where the ground state changes to some other state, which is the right memory state, assuming $\chi \not\in \mathcal M$ and enough precision.

\bibliography{literature}

\end{document}